\documentstyle[12pt]{article}
\def\ddot{\mathaccent"707F}

\title{Inflatonless inflation}
\author{{\large \bf A. Dobado and A. L\'opez} \\
Departamento de F\'{\i}sica Te\'orica \\
Universidad Complutense de Madrid\\
 28040 Madrid, Spain }
\begin{document}
\maketitle
\begin{abstract}
Whichever could be the real theory of gravitation, the corresponding
low-energy effective lagrangian  will probably contain
higher
derivative terms. In this work we study the general conditions on
those terms
in order to produce enough inflation to
solve some of the problems of the standard
 Friedmann-Robertson-Walker cosmology in absence of any inflaton field.
We apply our results to some particular scenarios where higher
derivative terms appear in the effective lagrangian for gravity like
those coming from graviton (two)-loops or integrating out ordinary matter
(like the one present in the Standard Model) .
 \end{abstract}
\vskip 3.0cm
PACS: 98.80.Cq \\
FT/UCM/13/94

\newpage
\baselineskip 0.83 true cm
\newpage
\section{Introduction}

In the last years, the general physical idea of a period of enormous
inflation in the early history of our  universe has emerged as one of the
most simple and successful ways to solve many of the problems of the
Friedmann-Robertson-Walker cosmological scenario [1,2]. The list of those
problems includes flatness, the horizon problem and the origin of
the density fluctuations needed to produce the current galactic structure.

Concerning to the nature of the physical mechanism resposible for the
inflation there are many of them proposed in the literature. However no one
has universal acceptance. Typically they include the action of some scalar
field called inflaton taking some  vacuum expectation value. The corresponding
energy density plays the role of an effective cosmological constant which
gives rise to a de Sitter phase of exponential expansion. The concrete nature
of the inflaton field depends on the different microphysics models considered.

In this paper we deal with another kind of approach based on the possibility
of having inflation without any inflaton field.
This possibility was envisaged by Starobinsky some time ago
[3]. Even before the inflationary paradigm was established, this author
discovered that the addition of  some  terms to the Einstein
equation of motion gives rise to de Sitter spaces as solutions. In fact those
terms can be obtained in the effective action for gravity as the result of
integrating out  conformal free matter fields.
The possibilities for the Starobinsky mechanism to produce successful
inflation were studied by Starobinsky himself and by Vilenkin in [3,4].
More recently it was realized by the authors [5]
 that the standard Einstein action
supplemented with a six derivative term, gives rise to  a modified equation
of motion supporting de Sitter solutions. Moreover, the introduction of this
term is not arbitrary at all but it is necessary for the renormalizability of
the effective low-energy theory of gravitation at the two-loop level [6].

Whatever could be the most fundamental theory of gravitation (like superstrings
or any other) it is clear that at low energy it must lead to the standard
Einstein lagrangian proportional to the scalar curvature $R$ which has two
derivatives of the metric $g_{\alpha\beta}$ and therefore leads to graviton
scattering amplitudes of the order of the external momenta over the Plank mass
$M_P$ squared.

This is a good
approximation at low energies but when higher energies are considered, higher
derivative terms will in general appear in the effective gravitational
lagrangian. These terms will be affected by adimensional constants and the
necessary $M_P$ factors depending on the dimension of the operator. The
adimensional constants play a double role: First they carry the information
about the underlying theory of gravitation, second they will absorb the
divergences that appear when quantum corrections i.e. loops are computed with
the effective lagrangian. These loops can contain matter fields, like in the
Starobinsky case, or gravitons (toghether with the corresponding ghosts)
as it happens in the model considered in [5] or whatever.
This scheme has many analogies with the
phenomenological lagrangian approach proposed  by Weinberg [7] for the
description of the low-energy hadron interactions and  further elaborated by
Gasser and Leutwyler [8]. This technic has also been applied to
the parametrization of the scattering amplitudes of the longitudinal
components of the electroweak bosons at the $Tev$ scale [9]. More recently
[5,10] the same phylosophy has been proposed to describe the low-energy action
for gravity [11].

At this point a natural question arises: Which are the new higher derivative
terms that one must include in the low-energy effective pure gravity
lagrangian?
In principle, any general covariant combination of scalar curvature and the
Ricci
and Riemann tensors $R, R_{\alpha\beta}$ and $R_{\alpha\beta\gamma\delta}$
should be included.  For example one has three possible independent four
derivative terms that can be written as $R^2$,
 $R^{\alpha\beta}R_{\alpha\beta}$ and
$-4R^{\alpha\beta}_{\;\;\;\;\gamma\delta}R^{\gamma\delta}_{\;\;\;\;\alpha\beta}
+16R^{\alpha\beta}R_{\alpha\beta}-4R^2$. The third
term  is a total derivative  related
with the Euler class of the
space-time manifold.

 In the general case we will assume the action for
gravity to be the space-time integral of an arbitrary local analytical scalar
function of the  scalar curvature and the Ricci and Riemann tensors added
to the standard Hilbert-Einstein action or, in other words:
\begin{equation}
S_G=\int d^nx \sqrt{-g}(-\frac{R+\lambda}{16\pi \bar G}+F(R,R_{\alpha\beta},
R_{\alpha\beta\gamma\delta}))
\end{equation}

where $\bar G$ denotes the gravitational constant in $n$ dimensions. Note
however
that in principle the effective action for gravity should contain also
non-local
terms but these  will not be considered here.

As it was discussed above, in this work we are
interested in studying the possibilities of having inflation without
the introduction of any inflaton field. Therefore we have to deal
with the problem of the  determination of the precise conditions
to be required on the function $F$ appearing in the action in  eq.1 leading to
de  Sitter  space-times as solutions of the corresponding equation of motion.
\newline
\\

\section{Existence of (anti-)de Sitter solutions}

For maximally symmetric space-times it is always possible to write the Riemann
tensor just in terms of the scalar curvature:
\begin{equation}
R_{\alpha\beta\gamma\delta}=\frac{1}{n(n-1)}R(g_{\alpha\gamma}g_{\beta\delta}
-g_{\alpha\delta}g_{\beta\gamma})
\end{equation}
where $n$ is the dimension of space-time.
So, we can rewrite the action integral in eq.1 (in four dimensions without
cosmological constant) in terms of the scalar curvature only:

\begin{equation}
S_G=\int d^4x \sqrt{-g}(-\frac{M_p^2}{16\pi}R+G(R))
\end{equation}

for some well defined analytical function $G$ to be obtained from the original
$F$ function in eq.1. We will first do the study in four dimensions but at the
end of this section we will consider also more general scenarios.

 The equation of motion corresponding to this action
 can be found to be:
\begin{equation}
\frac{{M_p}^2}{16\pi}(R_{\alpha\beta}-\frac{1}{2}g_{\alpha\beta}R)=
-\frac{1}{2}g_{\alpha\beta}G(R)+R_{\alpha\beta}G'(R)+g_{\alpha\beta}\Box
G'(R)-G' (R)_{;\alpha\beta}
\end{equation}

where a prime denotes derivative respect to $R$.
In order to find the condition for having inflationary space-times as solution
of this equation it is important to remember that the (anti-)de Sitter
space-time is a maximally symmetric space of constant scalar curvature $R$.
Therefore,  for this case the above equation of motion reduces to:

\begin{equation}
\frac{{M_p}^2}{16\pi}R=2G(R)-RG'(R)
\end{equation}
which is the condition for having (anti-)de Sitter solutions. In principle,
this equation can have zero, one or several solutions. In the first case no
inflation is  possible but in the other cases one or more inflationary
phases can  be present. It is immediate to see from the equation above  that
the addition of quadratic terms to the Einstein-Hilbert action (i.e. $F=\alpha
R^2+ \beta R_{\mu\nu}R^{\mu\nu}+\gamma R_{\alpha\beta\gamma\delta}
R^{\alpha\beta\gamma\delta}$, or in other words $G(R)\propto R^2$) does not
drive to any solutions different from $R=0$ which corresponds to the
Minkowsky space.

Now let us extend this analysis to arbitrary dimensions. The interest in
studying higher derivative gravitational actions in higher dimensions arises
from the fact that they could provide a mechanism for obtaining the
compactification of the extra dimensions in theories like Kaluza-Klein or
superstrings [12],[13], on the other hand it has been also shown that this
kind of actions drive to modified Einstein equations for the ordinary four
dimensional space-time supporting inflationary solutions [14].

Following the previous study in four dimensions, we will consider now the
$n=4+D$ dimensional action in eq.1 and we will concentrate in studying the
conditions on function $F$ that lead to a de Sitter geometry for the four
dimensional space-time and to spontaneous compactification of the extra
dimensions, or in other words, we search under which conditions the equations
of motion derived from this action admit a solution corresponding to a product
space-time $M\times K$ with $M$ a maximally symmetric four dimensional
space-time
and for simplicity we will asumme $K$ to be a compact D-sphere $S^D$.
Therefore we can write the $4+D$ dimensional metric tensor as follows:

\begin{eqnarray}
d\tau
^2=-g^M_{\mu\nu}dx^{\mu}dx^{\nu}-g^K_{ij}dy^idy^j
\end{eqnarray}

with $g^K_{ij}dy^idy^j=b^2(t)d\Omega_D ^2$ and where we have splitted the
coordinates on $M\times K$ as $X=(x^{\mu},y^j)$, $x^{\mu}$ with $\mu=1,..,4$
space-time coordinates, $y^j$ with $j=5,..,4+D$ internal coordinates and
$d\Omega_D ^2$ is the metric tensor on a unit D-sphere. It is important to
notice that the temporal dependence of $b(t)$  would appear in the four
dimensional effective action as a temporal dependence of the physical
constants ($G$, $\lambda$ and the coefficients of the higher order terms),
since strong  limits [15]  have been imposed to the range of variability of
fundamental constants throughout the evolution of the Universe, it is in
principle quite accurate to make $\dot b=0$. Taking this into account, any
scalar contraction of a $4+D$ dimensional tensor can be written as a sum of
the corresponding four and $D$ dimensional contractions, for instance: $\left.
R^{\mu\nu}R_{\mu\nu}\right|_{M\times K}=\left.
R^{\mu\nu}R_{\mu\nu}\right|_{M}+\left. R^{\mu\nu}R_{\mu\nu}\right|_{K}$, this
fact together with eq.2 enables to write the action integral as follows:

\begin{eqnarray}
S=\int d^4 x d^D y \sqrt{-g_M} \sqrt{-g_K} (
-\frac{R_M}{16\pi \bar G}-\frac{R_K}{16\pi \bar G} \\ \nonumber
-\frac{\lambda}{16\pi \bar
G}+ G_4(R_M)+G_D(R_K)+H(R_M,R_K))
\end{eqnarray}

Here we have formally gathered terms depending only on $R_M$ in function
$G_4$, terms depending on $R_K$ in $G_D$, crossed terms are included in
$H(R_M,R_K)$ and we assume without loosing
generality $G_D(0)=G_4(0)=H(0,R_K)=H(R_M,0)=0$.

The equations of motion corresponding to this action can be found in a similar
fashion to eq.4 and using the fact that $M$ and $K$ are both spaces of constant
curvature they drive to:

\begin{eqnarray}
\frac {1}{16\pi\bar G}(-\frac {1}{4}R_M -\frac {1}{2}(R_K+\lambda -16\pi\bar G
G_D(R_K)))= \\ \nonumber
-\frac {1}{2}(G_4(R_M)+H(R_M,R_K))+\frac
{R_M}{4}(G_4'(R_M)+\frac{\partial H}{\partial R_M})
\end{eqnarray}

for the ordinary space-time metric components and for the internal components
we
obtain:

\begin{eqnarray}
\frac {1}{16\pi\bar G}(\frac {2-D}{2D}R_K -\frac {1}{2}(R_M+\lambda -16\pi\bar
G
G_4(R_M)))= \\ \nonumber
-\frac {1}{2}(G_D(R_K)+H(R_M,R_K))+\frac
{R_K}{D}(G_D'(R_K)+\frac{\partial H}{\partial R_K})
\end{eqnarray}

These are the conditions, analogous to eq.5, that the function $F$ should
satisfy in order to get the compactification of the internal dimensions
provided the ordinary space-time is a de Sitter one with constant scalar
curvature $R_M$. In principle these are rather natural conditions and no fine
tunning on $F$ seems to be needed to satisfy them.  However we are interested
only in solutions of eq.8 and eq.9 leading to solutions  of the standard
Einstein equations for the external space in the limit of low external
curvature. Then, in that limit we obtain the following two conditions on
$G_D(R_K)$:
\begin{equation}
G_D(R_K)=\frac {R_K+\lambda}{16\pi \bar G}
\end{equation}
and

\begin{equation}
G'_D(R_K)=\frac {1}{16\pi \bar G}
\end{equation}

It is then possible to conclude that for an arbitrary function $G_D(R_K)$ the
above two conditions will not be simultaneously fulfilled in general and
certain
unnatural choices of the coefficients of $F$ will be needed. This is in
contrast with the ordinary case where  a Minkowsky $R_M=0$ solution
always exits in eq.5. In particular we have checked that for the
higher order term studied in [13]:
\begin{equation}
F=\alpha
R^2+ \beta R_{\mu\nu}R^{\mu\nu}+\gamma R_{\alpha\beta\gamma\delta}
R^{\alpha\beta\gamma\delta}
\end{equation}

the conditions in eq.10 and eq.11 drive to the fine tunning of the cosmological
constant found there.

In principle this results strongly depend on the topology of the
compactified dimensions, however it is possible to show that, in the case of
a static D-torus and a de Sitter external space-time, one is faced with the
same problem, namely, the fine tunning of the parameters, specifically the
equations of motion drive in this case to:

\begin{equation}
G_4(R_M)=\frac {R_M+\lambda}{16\pi \bar G}
\end{equation}

and

\begin{equation}
G'_4(R_M)=\frac {1}{16\pi \bar G}
\end{equation}
Therefore, everything seems to indicate that, the higher dimensional
context,
requires the fine tunning of the parameters of the initial action in order to
have, at least asymptotically, an internal space of constant size and
simultaneously recover the Einstein equations of motion for the external
space in the limit of low external curvature. For this reason, in the following
we will concentrate only  in the four dimensional case.
  \newline \\

\section{Robertson-Walker perturbations}

 Let us consider now the standard Robertson-Walker metric:

\begin{equation}
d\tau ^2=dt ^2-a^2(t)(\frac{dr^2}{1-kr^2}+r^2d\theta^2+r^2sin^2\theta d\phi^2)
\end{equation}
(we will follow the notation convention of Weinberg's book [16]).

where $k=1,0,-1$ corresponds to a closed, flat or open space and $a(t)$
is the universe scale parameter. For the sake of simplicity, in the following
we will concentrate in the flat or $k=0$ universe. This is not in general a
maximally symmetric space-time but it includes a particular case which indeed
is, namely, $a(t)=a(t_0)\exp H_0(t-t_0)$. This fact enables to use this metric
to study small perturbations around the (anti-)de Sitter solutions found in
eq.5  and, in turn, the perturbative analysis will provide an estimation of the
duration of each inflationary period as we will see later on. Proceeding in
this way  we first note that the Riemann tensor for this metric has only the
following six non-vanishing independent components:

\begin{equation}
R^{tr}_{\;\;\;\;tr}=R^{t\theta}_{\;\;\;\;t\theta}=R^{t\phi}_{\;\;\;\;
t\phi}=-\frac{\ddot {a}}{a}
\end{equation}
\begin{eqnarray}
R^{\theta\phi}_{\;\;\;\;\theta\phi}=R^{r\theta}_{\;\;\;\;r\theta}=R^{r\phi}
_{\;\;\;\;r\phi}=-\frac{{\dot{a}}^2}{a^2}
\end{eqnarray}

 Since only this functions of $a$ and its derivatives will appear in the
action it is then more useful in practice to work with them as new variables,
thus we define: $b(t)=log(a(t))$, $H(t) \equiv \dot
b(t)=\dot {a}/a$ and $\dot H(t)=\ddot b(t)=\ddot {a}/a -
\dot {a}^2/a^2$. In terms of the Hubble parameter H(t) that we have just
defined, the (anti-)de Sitter space corresponding to $a(t)=a(t_0)\exp
H_0(t-t_0)$ is simply $H=H_0$ and the action integral in eq.1 reads:

\begin{equation}
S_G\propto \int dt e^{3b} L(H,\dot H)
\end{equation}

here a global volume factor has been extracted and the function $L(H,\dot H)$
which is obtained from the function $F$ in eq. 1 includes the Einstein-Hilbert
part of the lagrangian. The equation of motion for the Hubble parameter $H$
obtained from the action above reads:

\begin{equation}
3L -3H \frac{\partial L}{\partial H}-\frac{d}{dt}\frac {\partial L}{\partial
H} + 3\dot H \frac {\partial L}{\partial\dot H}+9 H^2 \frac {\partial L}
{\partial\dot H}+6H\frac {d}{dt}\frac{\partial L}{\partial\dot H}+\frac
{d^2}{dt^2} \frac{\partial L}{\partial\dot H}=0
\end{equation}

This equation obviously posseses the solutions $H(t)=H_0$ corresponding
to the (anti-)de Sitter space found in eq. 5. In order to study the stability
of these solutions we must consider the behavior of small perturbation around
them, that is:

\begin{equation}
H(t)=H_0+\delta (t),\;
 \dot H(t)=\dot \delta (t)
\end{equation}

Expanding $L(H,\dot H)$ and its derivatives to the first order in $\delta$ we
find:

\begin{equation}
L=L_0+L_1\delta+L_2 \dot\delta +O(\delta^2)
\end{equation}
\begin{equation}
\frac{\partial L}{\partial H}=L_1+L_{11}\delta+L_{12}\dot\delta+O(\delta^2)
\end{equation}
\begin{equation}
\frac{\partial L}{\partial \dot
H}=L_2+L_{21}\delta+L_{22}\dot\delta+O(\delta^2)
\end{equation}

where we have defined the following coefficients:

\begin{eqnarray}
L_0=L(H_0,0) \nonumber \\
 L_1=\frac{\partial L}{\partial H} (H_0,0)\nonumber \\
L_{2}=\frac{\partial L}{\partial \dot H} (H_0,0)  \nonumber \\
L_{11}=\frac{\partial^2 L}{\partial H^2} (H_0,0)\nonumber \\
L_{12}=L_{21}=\frac{\partial^2 L}{\partial H\partial \dot H} (H_0,0) \nonumber
\\
L_{22}=\frac{\partial^2 L}{\partial \dot H^2} (H_0,0) \end{eqnarray}

Finally, substituting these expansions in eq.19 and neglecting terms of
second order we obtain the linearized equation for $\delta (t)$, which can be
written in the following way:

\begin{equation}
3(L_0-H_0L_1+3{H_0}^2L_2)+\frac {1}{3H_0} \frac{d}{dt} {\cal F}+{\cal F}=0
\end{equation}

where $\cal F$ stands for:

\begin{equation}
{\cal F}=L_{22}\ddot\delta+3H_0L_{22}\dot\delta+\delta(6L_2
+3H_0L_{12}-L_{11})
\end{equation}

Now the condition in eq. 5 for having (anti-)de Sitter
solutions can be rewritten as a condition on $H_0$:

\begin{equation}
L_0-H_0L_1+3{H_0}^2L_2=0
\end{equation}

The resolution of the linearized equation for $\delta (t)$ (eq.25) around
each solution ${H_0}^{(i)}$ of eq.27 simply becomes the resolution of equation
${\cal F}=0$ whose solutions are  any linear combination of modes $exp
(t/\tau)$ with:

\begin{equation}
\frac{1}{\tau^{(i)}}=-\frac{3H_0^{(i)}}{2}\pm
(\frac{9{{H_0}^{(i)}}^2}{4}-\frac{1}{L_{22}^{(i)}}(6L_2^{(i)}+
3H_0^{(i)}L_{12}^{(i)}-L_{11}^{(i)}))^{(1/2)}
\end {equation}

Here $i$ runs over the number of real solutions of eq.27.
The stability of this inflationary solutions is given by the sign of
$\tau^{(i)}$, if $\tau^{(i)}<0$ then the corresponding  solution is stable,
otherwise it is unstable. From eq.28 an stable mode is always present. The
value of $\tau^{(i)}$ for unstable modes gives an estimation of the duration
of the corresponding inflationary period and, in turn, of the number of
$e$-folds $N_e^{(i)}$ produced during this period:

\begin{equation}
N_e^{(i)}=\int H(t) dt=H_0^{(i)} \tau^{(i)}
\end{equation}

With the simple method described above it becomes very easy
to decide if some given generalized effective action for gravity
gives rise or not to exponential inflationary solutions and, in that case,
to obtain an estimation  of the change of the scale produced in the
coresponding inflationary phase. In the next sections we will apply the method
to some different scenarios which have been considered in the literature in
different contexts.
\newline
\\

\section{The two-loop counterterm}

The first example we will consider is that of [5]. There the authors
studied the minimal consistent effective low energy two-loop renormalizable
lagrangian for pure gravity, this lagrangian is shown to contain a six
derivative term [6] added  to the usual Einstein-Hilbert one:

\begin{equation}
{\cal L}_{eff}=-\frac {{M_p}^2}{16\pi}R+\frac {\alpha}{{M_p}^2}
R^{\alpha\beta}_{\;\;\;\;\delta\gamma}R^{\delta\gamma}_{\;\;\;\;\sigma\rho}
R^{\sigma\rho}_{\;\;\;\;\alpha\beta}
\end{equation}

The function $L(H,\dot H)$ appearing in eq.18 is obtained just by rewritting
eq.30 for the flat Robertson-Walker space:

\begin{equation}
L(H,\dot H)=\frac {6{M_p}^2}{16\pi}(\dot H +2H^2)-24 \frac
{\alpha}{{M_p}^2}((\dot H+H^2)^3+H^6)
\end{equation}

By solving the eq.27 we find the inflationary periods produced as a
consequence of the introduction of this higher order term. The equation
reads:

\begin{equation}
\frac {18{M_p}^2}{16\pi} {H_0}^2+\frac {72\alpha}{{M_p}^2}{H_0}^6=0
\end{equation}

with the obvious solutions $H_0=0$ and
${H_0}^4=-{M_p}^4/(64\pi\alpha)$. Therefore, there exists only one inflationary
period provided $\alpha$ is negative. On the other hand, eq.29 gives the number
of $e$-folds produced before the Universe leaves this stage of exponential
growing, it is immediate to obtain: $N_e\simeq 4.81$. It is important to notice
that $N_e$ does not  depend on the $\alpha$
coefficient preceding it. Therefore, the addition of the six derivative term
to the Einstein action considered here does not seem to produce
enough inflation to solve the problems of the standard cosmology.
 \\

\section{The effect of  matter}

 This model is based on the
semiclassical Einstein equations:

\begin{equation}
R_{\mu\nu}-\frac{1}{2} g_{\mu\nu} R=-8\pi G<T_{\mu\nu}>
\end{equation}

Where $<T_{\mu\nu}>$ is the vacuum expectation value of the stress tensor of a
number of massless
 conformally invariant quantum fields with different spin values. As it is
well known this vacuum expectation value will be divergent in general and some
regularizarion method will be needed to give it a sense. These divergences
affecting  $<T_{\mu\nu}>$ has been computed in the literature and cause the
appearance of fourth order operators at the leading adiabatic order (see [17]
for a very complete review). The corresponding higher order terms in the
matter  lagrangian  may be considered also as part of the gravitational
effective lagrangian since they are pure geometric objects i.e. they depend
only on the standard Riemann and Ricci tensor and the curvature scalar, but
not on the
matter fields themselves. In dimensional regularization and using the fact that
$-4R^{\alpha\beta}_{\;\;\;\;\gamma\delta}R^{\gamma\delta}_{\;\;\;\;\alpha\beta}
+16R^{\alpha\beta}R_{\alpha\beta}-4R^2$ is a total divergence in four
dimensions, this divergent  contribution to  the effective lagrangian can be
written for massless fields as:

\begin{equation}
{\cal L}_{div}=\frac {1}{{(4\pi)}^2}(\frac {1}{\epsilon} -\frac {\gamma_e}{2}
)a_2(x) \end{equation}
 with:
\begin{equation}
a_2(x)=\beta_1 R^2+\gamma R_{\mu\nu}R^{\mu\nu}
\end{equation}

\begin{equation}
\beta_1=-\frac {N_{sc}}{180}+\frac {N_{gh}}{90} + \frac {N_{\nu}}{60}+\frac
{N_{fe}}{30}-\frac {N_{v}}{20}
\end{equation}

\begin{equation}
\gamma=\frac {N_{sc}}{60} - \frac {N_{\nu}}{20}-\frac {N_{fe}}{10}+\frac
{7N_{v}}
{30}-\frac
{N_{gh}}{30}
\end{equation}
where we have assumed $N_{sc}$ scalars, $N_{gh}$ ghosts,$N_{\nu}$ neutrinos,
$N_{fe}$ Dirac fermions and $N_{v}$ vectors fields to be present.

Now we have to deal with the problem of the divergent coefficient multiplying
these terms. In principle, one may estimate in a heuristic way the value of the
corresponding renormalized coefficients by performing the substitution $ 1/
\epsilon -\gamma_e / 2 \rightarrow log(\frac {M_p}{M})$ which is equivalent to
the assumption of integrating the matter fields modes from some infrared cutoff
$M$ to the $M_p$ scale. Thus, the corresponding  renormalized
effective lagrangian will be:

\begin{equation}
{\cal L}_G=-\frac {{M_p}^2}{16\pi} R
+\frac {1}{{(4\pi)}^2}\log(\frac {M_p}{M})a_2(x)
\end{equation}

On the other hand, the finite remainder $<T_{\mu\nu}>_{ren}$  is in general
quite difficult to compute. However, in the case of conformally flat spaces
and quantum fields which are also conformally invariant $<T_{\mu\nu}>_{ren}$
can be exactly computed out of the knowledge of the trace anomaly and it
renders:

\begin{equation}
<T_{\mu\nu}>_{ren}=\beta_2 ^{(1)}H_{\mu\nu}+\rho
^{(3)}H_{\mu\nu}+^{(4)}H_{\mu\nu}
\end{equation}
where:
\begin{equation}
^{(1)}H_{\mu\nu}=2R_{;\mu\nu}-2g_{\mu\nu} \Box R -\frac {1}{2}
g_{\mu\nu}R^2+2RR_{\mu\nu}
\end{equation}

 \begin{equation}
^{(3)}H_{\mu\nu}=R_{\mu}^{\;\;\sigma}R_{\nu\sigma}-\frac{2}{3}RR_{\mu\nu}-\frac{1}{2}
g_{\mu\nu}R^{\sigma\tau}R_{\sigma\tau}+\frac{1}{4}g_{\mu\nu}R^2
\end{equation}
$^{(4)}H_{\mu\nu}$ is a border term and will be omitted. It is possible
to derive the first term above $^{(1)}H_{\mu\nu}$ from an action integral
in this way:
 \begin{equation}
^{(1)}H_{\mu\nu}=\frac {1}{\sqrt {-g}} \frac {\delta}{\delta g^{\mu\nu}}
\int \sqrt {-g} R^2 d^n x
\end{equation}

 Note that in general $^{(3)}H^{\mu\nu}$ cannot be obtained by varying a
local action, although several non-local actions have been proposed which
drive to this term. In spite of this, in the case we are considering,
conformally flat space, a local action can be found
[18] which can be written in terms of the Hubble parameter as:
\begin{equation}
\Gamma=\int dt e^{3b} H^4
\end{equation}
 It is then possible to write a matter lagrangian in the
case of conformally flat space which drives to $<T_{\mu\nu}>_{ren}$ and
in terms of the Hubble parameter takes the form :
\begin{equation}
 L_M=\beta_2 (\dot H+2H^2)^2+\rho H^4
\end{equation}

with the coefficients $\beta_2$ and $\rho$ given by:
\begin{equation} \beta_2=-\frac {1}{144\pi^2} \frac
{1}{120}(N_{sc}+3N_{\nu}+6N_{fe}-18N_{v}-2N_{gh})
\end{equation}
\begin{equation}
\rho=-\frac {1}{8\pi^2} \frac
{1}{360}(-N_{sc}-\frac {11}{2}N_{\nu}-11N_{fe}-62N_{v}+2N_{gh})
\end{equation}
Therefore we can write the function $L(H,\dot H)$ in eq.18 including the
 gravitational and matter sectors as: \\

 \begin{eqnarray}
L(H,\dot H)=L_G+L_M=\frac {6{M_p}^2}{16\pi}(\dot H +2H^2)+\rho H^4 \nonumber \\
+36(\frac{1}{(4\pi)^2} \log (\frac {M_p}{M})\beta_1 +\beta_2)(\dot H +
2H^2)^2 \nonumber \\
 + 12 \frac {1}{(4\pi)^2}\log (\frac {M_p}{M}) \gamma (\dot
H ^2+3\dot H H^2 +3H^4)
\end{eqnarray}

As we did in the previous example we use eq.27 in order to get the
inflationary periods and we obtain in this case:
\begin{equation}
-\frac {M_p^2}{8\pi}H_0^2+\rho H_0^4=0
\end{equation}
whose solutions are simply $H_0=0$ and $H_0^2=M_p^2/(8\pi\rho)$. Thus we find
again only one de Sitter phase provided $\rho$ is positive. Finally eq.29 will
provide the number of e-folds during this phase. For an infrared cut-off
$M\simeq 100 GeV$ eq.29 yields:
 \begin{equation}
\frac {1}{N_e}=-\frac {3}{2} +\sqrt{\frac {9}{4}+\frac
{6\rho}{8.92\beta_1+2.97\gamma+36\beta_2}}
\end{equation}
 For the Standard Model of elementary
particles interactions based on the
gauge group $SU(3)_C{\times}SU(2)_L{\times}U(1)_Y$ we have $N_{sc}=4$,
 $N_{\nu}=3$, $N_{fe}=21$, $N_v=12$ and $N_{gh}=12$. By subtituting for these
values in eqs.48,49, one obtains: $H_0=1.52M_p$ and $N_e\simeq 44$. In
a similar fashion, other models with additional content of matter fields
can be also considered.
\\

\section{Conclusions}

In this work we have studied the possibility of having a phase
of exponential inflation starting from an effective lagrangian for
gravity which is an arbitrary function of the  Riemann and Ricci tensors and
the curvature scalar. Therefore, our main assumption is that there is some
epoch in the early history of our universe where its evolution can be
described in a classical way but with an unknown effective
action for the gravitational field.
We have set the precise conditions on this action for such an exponential
expansion to take place. We have shown that it is a rather common
phenomenon in four dimensions,
even without adding a cosmological constant, since for a generic action
eq.5 may have solutions different from zero.

However, in higher dimensions the
appearance of additional conditions makes it neccesary a precise (fine)
tunning of some parameters of the action. Therefore, these models do not seem
to provide a natural mechanism for achieving inflation with modified
gravitational actions.

 We have also studied the stability of the inflationary phase in four
dimensions and estimate its duration in classical terms.  With these formal
tools we have considered some particular models giving rise to higher
derivative terms for the gravity action.

The first case to which we have applied our general method is the
study of the effect of including the six derivative term needed
to renormalize two-loop pure quantum gravity. With this new term added
to the standard Einstein action we find that an exponential inflationary
solution appear. The corresponding  inflationary phase produces $4.8$ foldings
independently of the concrete value of the six derivative term coupling. Thus
we can conclude that this new term does not seem
to produce inflation enough to solve any of the problems
of the standard cosmology.

As a second example we have applied our general  method to study
the posiblility of having inflation driven by the terms induced in the
effective low-energy action for gravitation by integrating out the matter
fields. In particular we have found
the  interesting result that the Standard Model matter produce, by
itself, an inflationary phase with $N_e\simeq 44$. This is probably not
sufficient to solve the flatness and the horizon problem (see
 [19] for a recent disscussion about the precise conditions needed for that)
but it is large enough to take seriously the possibility of having an
Standard Model driven larger inflation as an output of more detailed
computations.

In conclussion we consider that, in absence of a fundamental theory of
gravitation,
the possibility of having inflation without the somewhat artifitial artifact of
the inflaton field, should not be ruled out. Moreover, the results found here
for the case of the Standard Model matter coupled to classical gravity, seems
to suggest that it is
worth to study in deep the dynamics of this system that, finally, is the only
one that we are sure is realized in Nature. Work is in progress in this
direction.

 {\bf Aknowledgments:}
This work has been partially supported by the Ministerio de Educaci\'on y
Ciencia (Spain)(CICYT AEN90-0034).

{\newpage}

\end{document}